\documentclass[fleqn,usenatbib]{mnras}

\usepackage{newtxtext,newtxmath}

\usepackage[T1]{fontenc}
\usepackage{ae,aecompl}

\usepackage{subcaption}

\usepackage{xcolor}
\usepackage{graphicx}	
\usepackage{amsmath}	

\usepackage{epsfig}
\usepackage{gensymb}
\usepackage{rotating}
\usepackage{color}
\usepackage{tabularx}
\usepackage{multirow}
\usepackage{graphics}
\usepackage{epsfig}
\usepackage{epstopdf}
\usepackage[para,online,flushleft]{threeparttable}
\usepackage[ampersand]{easylist}
\usepackage{hyperref}
\hypersetup{
    colorlinks=true,
    linkcolor=blue,
    filecolor=blue,      
    urlcolor=blue,
    citecolor=blue,
    filecolor=blue,
}
\title[Optical observations of G206.7+5.9]{Optical investigation of supernova remnant G206.7+5.9}

\author[H.~Bak{\i}\c{s} et al.]{{H.~Bak{\i}\c{s}$^{1}$,
E.~Aktekin$^{2}$\thanks{{\color {blue}corresponding author: ebrucaliskan@sdu.edu.tr} (EA)},
V.~Bak{\i}\c{s}$^{1}$,
H.~Sano$^{3}$ and
A.~Sezer$^{4}$}\\
$^{1}$Department of Space Sciences and Technologies, Akdeniz University, Antalya 07058, Turkey\\
$^{2}$Department of Physics, S\"{u}leyman Demirel University, Isparta 32000, Turkey \\
$^{3}$Faculty of Engineering, Gifu University, 1-1 Yanagido, Gifu 501-1193, Japan\\
$^{4}$Department of Computer Engineering, Avrasya University, Trabzon 61250, Turkey\\
}

\date{Accepted XXX. Received YYY; in original form ZZZ}

\pubyear{2025}

\begin{document}
\label{firstpage}
\pagerange{\pageref{firstpage}--\pageref{lastpage}}
\maketitle

\begin{abstract}
The shell-type supernova remnant (SNR) G206.7+5.9 was recently discovered in the radio band with the Five-hundred-meter Aperture Spherical radio Telescope (FAST). The remnant spans about 3$\fdg$5 in diameter and exhibits bilateral shells. In this work, we present optical spectra of G206.7+5.9 with  the Large sky Area Multi-Object fiber Spectroscopic Telescope (LAMOST), and narrow-band (H$\alpha$ and [S\,{\sc ii}]) images with the 1-m T100 telescope. The filamentary structure seen in H$\alpha$ image shows a clear correlation with the radio emission. We use optical line ratios to determine the physical parameters of G206.7+5.9. The LAMOST spectra reveal large ratios of [S\,{\sc ii}]/H$\alpha$ $\sim$ (0.61$-$1.78) and [N\,{\sc ii}]/H$\alpha$ $\sim$ (0.63$-$1.92) consistent with that expected for a shock-heated SNR. The emission lines [O\,{\sc i}] $\lambda$6300, $\lambda$6363 detected in the spectra also support the presence of shocks. Electron density ($n_{\rm e}$) measurements based on the [S\,{\sc ii}] $\lambda$6716/$\lambda$6731 ratio suggest densities between 117 and 597~cm$^{-3}$. We estimate the pre-shock cloud density ($n_{\rm c}$) to be approximately 2.6$-$13.3~cm$^{-3}$. We also investigate the archival H{\sc i} data and have newly identified an expanding gas motion of the H{\sc i}, whose velocity span is approximately 10 km s$^{-1}$. We conclude that G206.7+5.9 is an SNR exhibiting properties remarkably similar to those seen in Galactic SNRs.
\end{abstract}




\section{Introduction}
Galactic supernova remnants (SNRs) are initially discovered through radio observations. SNRs predominantly emit non-thermal continuum radiation in the radio band, with typical spectral indices of $\alpha\sim -0.5$ (defined by $S$ $\propto$ $\nu ^{\alpha}$) (e.g. \citealt{Du17}). Currently, there are over 300 confirmed Galactic SNRs listed in catalogs (e.g. \citealt{Fe12, Gr24}). 

Many SNRs have been extensively studied at different wavelengths (e.g. \citealt{Ba16, Te17, Ko17, Fi23, Ar24, Li24}), providing valuable insights into the physical processes occurring throughout their evolution. A more comprehensive view of the nature and evolution of SNRs is essential for understanding their interaction with the interstellar medium (ISM). For instance, optical filaments form in cooling regions behind the shock front and indicate the presence of shock-heated gas. However, only 30 per cent of Galactic SNRs exhibit optical emissions, typically identified through narrow-band filter imaging surveys (e.g. \citealt{Ma02, Ne17, Pa22, Ba23, Pa24, Fe24}).

G206.7+5.9 was identified as an SNR according to its non-thermal nature by showing steep radio continuum spectra with the Five-hundred-meter Aperture Spherical radio Telescope (FAST) L-band observations and the Effelsberg $\lambda$11~cm polarization measurements \citep{Ga22}. The observations showed that G206.7+5.9 has a size of about 3$\fdg$5 in diameter with bilateral shells. It contains one shell on the eastern side and two shells on the western side. Using H\,{\sc i} observations, they determined an average value of $V_{\rm LSR}$ = 4.6~km~s$^{-1}$, indicating a kinematic distance of about 0.44 kpc. This situates G206.7+5.9 within the Local Arm, with an estimated physical size of approximately 27~pc.

The physical properties of G206.7+5.9 and its surrounding environment have not been studied in the other wavelengths. In this work, we investigate optical properties of the south-west (SW) and  north-west (NW) parts of G206.7+5.9 using the Large sky Area Multi-Object fiber Spectroscopic Telescope (LAMOST). We also obtain H$\alpha$ and [S\,{\sc ii}] images of the SW region using the 1-m T100 telescope of T\"{U}B\.{I}TAK National Observatory (TUG)\footnote{\url{https://tug.tubitak.gov.tr}} in T\"{u}rkiye. In the following, Section~\ref{obs} outlines the optical observations and data reduction. Section~\ref{analysis} presents the analysis and results. In Section~\ref{discuss}, we discuss the observational findings, and Section~\ref{conc} offers our conclusions.

\section{Optical observations and data reduction}

In this study, we conducted a detailed optical analysis of SNR G206.7+5.9, utilizing data from both LAMOST and the T100 telescope. Our approach integrates medium-resolution optical spectra from LAMOST's Data Release 9 (DR9) and complementary observations from T100, providing a unique dataset for analyzing the emission lines and physical properties of SNR regions.

Our work emphasizes optical diagnostics through the detection of faint emission lines. This optical analysis allows for the derivation of crucial physical parameters such as electron density and ionization states with greater precision than past efforts limited to a single instrument or wavelength range.

\label{obs}
\subsection{LAMOST data}
LAMOST is a quasi-meridian reflecting Schmidt telescope with an effective aperture of approximately 4 meters and a 5$\degree$ diameter field of view (FoV) \citep{Cu12}. It functions in two main observational modes: one is a low-resolution mode, and the other is a medium-resolution mode. 

The medium-resolution LAMOST spectra are divided into two channels: the blue channel covers wavelengths from 3700 to 5900 {\AA}, and the red channel ranges from 5700 to 9000 {\AA}. The slit (2/3) has a length of 144~mm and a width of 0.22~mm. 

We obtain the processed LAMOST data from the DR9 toward G206.7+5.9. The observations were conducted on November 24, 2020, with an exposure time of 900s and the code \texttt {pylamost}\footnote{\texttt {pylamost} is a python interface for accessing these data, supporting both low and medium resolution data queries and file downloads. The main code is \texttt {pylamost.py}, with sample code available in \texttt {pylamost-test.ipynb}. See \url{https://github.com/fandongwei/pylamost}} has been used to access the LAMOST spectral data.

\subsection{T100 data}
We performed H$\alpha$ and [S\,{\sc ii}] imaging of the SW region centred at RA=06$^{\rm h}$ 56$^{\rm m}$ 43$\fs$67; Dec=+06$\degr$ 44$\arcmin$ 27$\farcs$06 of G206.7+5.9 with the 1.0~m fully automatic Ritchey-Chrétien T100 telescope on September 27, 2024. The CCD camera consists of $4096\times4096$ pixels, each measuring 15 ${\mu}$m $\times$ 15 ${\mu}$m, covering a $21.5 \times 21.5$ arcmin$^2$ FoV. The details of the imaging observations are presented in Table~\ref{Table1}. We reduced the imaging data  with \texttt{Image Reduction and Analysis Facility} (\texttt{IRAF})\footnote{\url{https://iraf-community.github.io/}}. We began by subtracting the bias and applying flat-field corrections. We cleaned the bad pixels and removed the cosmic rays from the CCD images. Finally, we combined the images and then subtracted the continuum image from the combined image.

\begin{table*}
 \caption{Imaging observations of the SW region of G206.7+5.9 with T100 telescope.}
 \begin{tabular}{@{}p{2.5cm}p{2.1cm}p{2.5cm}p{2.7cm}@{}}
 \hline
Filter       			& Wavelength    	 & FWHM   & Exposure time     \\ 
         			&    (nm)	             &   (nm)          &  (s) \\  
 \hline
H$\alpha$     		    &     656.0             &   10.8           & $3 \times 1200$  \\
$[$S$\,${\sc ii}$]$       &     674.0             &   11.2           &     $3 \times 1200$  \\
Cont.  	    	        &     551.0             &   88     &   $3 \times 300$        \\
\hline
\label{Table1}
\end{tabular}
\end{table*}

\section{Analysis and Results}
\label{analysis}
\subsection{Optical images}
\begin{figure*}
\includegraphics[angle=0, width=7.5cm]{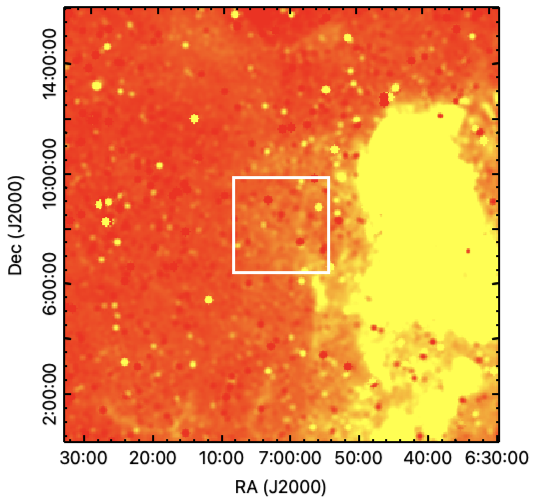}
\includegraphics[angle=0, width=7.2cm]{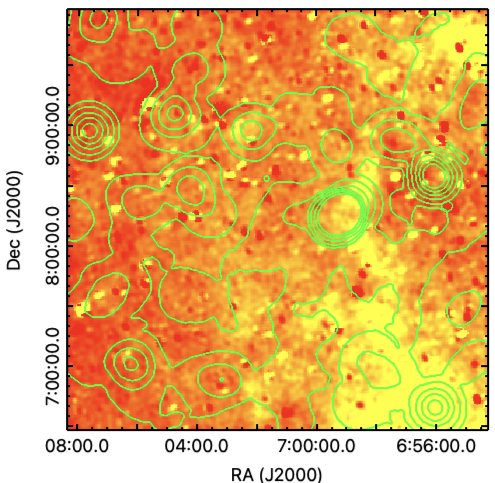}
\includegraphics[angle=0, width=7.5cm]{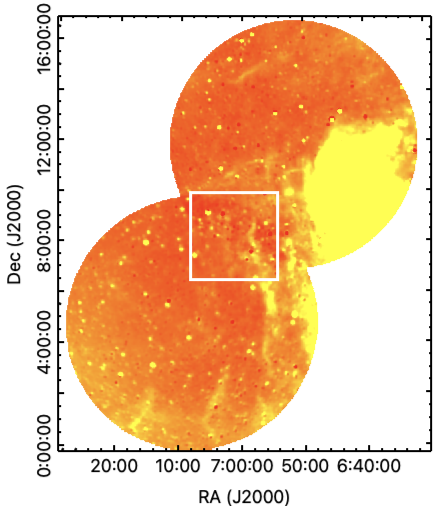}
\caption{H$\alpha$ images of G206.7+5.9 and its surroundings from two different surveys. Top panel: SHASSA \citep{Ga01} continuum-corrected H$\alpha$ image (left). The white box represents the entire SNR (3.5 $\times$ 3.5 deg$^{2}$, RA=07$^{\rmn{h}}$ 01$^{\rmn{m}}$ 18$\fs$8; Dec= 08$\degr$ 13$\arcmin$ 11$\farcs$6) given by \citet{Ga22}. The area shown in the white box is zoomed-in on the right side overlaid with the radio-continuum contours (green) at 1.4 GHz taken from the NRAO VLA Sky Survey (NVSS; \citealt{Co98}). The radio-continuum contour levels range from $-0.55$ to 1.11~mJy~beam$^{-1}$. Bottom panel: VTSS \citep{De98, Fi03} H$\alpha$ image. These images show a large number of filaments along the entire remnant.} 
\label{figure1a}
\end{figure*}

The SNR G206.7+5.9 has a large size in the radio continuum image from FAST given by \citet{Ga22}. To view the entire SNR and its surroundings, we present a map of H$\alpha$ emission from the Southern H$\alpha$ Sky Survey Atlas (SHASSA; \citealt{Ga01}) and Virginia Tech Spectral-line Survey (VTSS; \citealt{De98, Fi03}) in Fig. \ref{figure1a}.

\begin{figure*}
\includegraphics[angle=0, width=8.8cm]{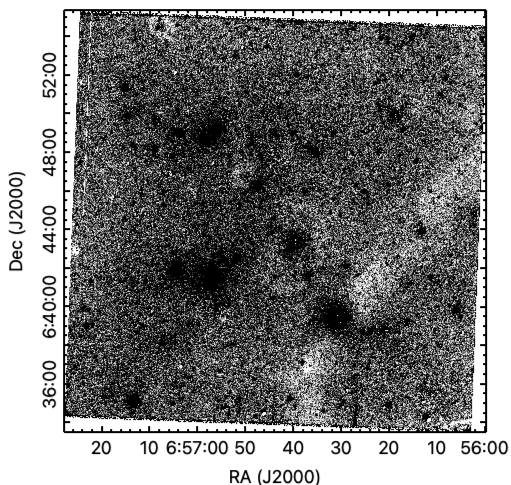} 
\includegraphics[angle=0, width=8.8cm]{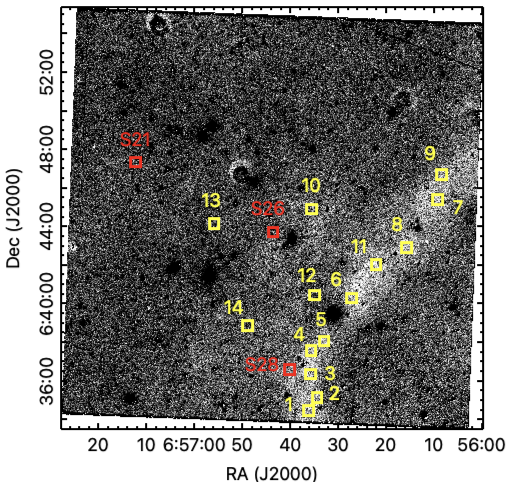}
\caption{Left panel: The H$\alpha$ (continuum-subtracted) image of the SW region with T100 telescope. Right panel: The [S\,{\sc ii}]/H$\alpha$ image with marked selected positions to calculate ratio values. Three of these positions, specifically S21, S26, and S28, align with the regions where we collected LAMOST spectra, marked in red on the figure.}
\label{figure1b}
\end{figure*}

We show the H$\alpha$ image of the SW region with T100 telescope in the left panel of Fig.~\ref{figure1b}. We provide the [S\,{\sc ii}]/H$\alpha$ image, generated using the \texttt {imarith} command in \texttt{IRAF}, which is displayed in the right panel of Fig.~\ref{figure1b}. Estimated [S\,{\sc ii}]/H$\alpha$ ratios and their errors  taken from 17 different regions on the image are given in Table~\ref{Table2}. 

\begin{table}
 \caption{[S\,{\sc ii}]/H$\alpha$ ratios for SW region of G206.7+5.9 obtained with the [S\,{\sc ii}]/H$\alpha$ ratio image.}
 \label{Table2}
 \begin{tabular}{@{}p{1.2cm}p{3.4cm}p{2.1cm}@{}}
\\[0.05 ex]
\hline
 ID &  R.A. (J2000) ; Dec. (J2000)  &  Ratio     \\
  & (h m s)  ~~~~~~~~~~~ ($\degr$ $\arcmin$ $\arcsec$)  & \\
\hline
1 &  06 56 36.2 ; ~+06 34 26.2  & 0.71 $\pm$ 0.02   \\

2 &  06 56 34.4 ; ~+06 35 07.9 & 0.66 $\pm$ 0.02     \\

3 &  06 56 35.8 ; ~+06 36 20.3  & 0.79 $\pm$ 0.04    \\

4 &  06 56 35.7 ; ~+06 37 32.9  & 0.51 $\pm$ 0.01     \\

5 &  06 56 32.9 ; ~+06 38 03.3  & 0.65 $\pm$ 0.02    \\

6 & 06 56 27.2 ; ~+06 40 16.8  & 0.52 $\pm$ 0.03   \\

7 &  06 56 09.2 ; ~+06 45 23.6  &  0.84 $\pm$ 0.02     \\

8 &  06 56 15.7 ; ~+06 42 53.0 & 0.93 $\pm$ 0.04      \\

9 &  06 56 08.4 ; ~+06 46 41.7   & 0.59 $\pm$ 0.02      \\

10 & 06 56 35.6 ; ~+06 44 53.5  & 0.73 $\pm$ 0.02      \\

11 & 06 56 22.1 ; ~+06 42 01.5  & 0.58 $\pm$ 0.01      \\

12 & 06 56 35.0 ; ~+06 40 26.2   & 0.72 $\pm$ 0.04      \\

13 & 06 56 55.9 ; ~+06 44 08.8   & 0.84 $\pm$ 0.05      \\

14 & 06 56 48.8 ; ~+06 38 52.1   & 0.67 $\pm$ 0.02      \\

S21 & 06 57 12.3 ; ~+06 47 19.4  & 0.70 $\pm$ 0.03      \\

S26 & 06 56 43.6 ; ~+06 43 42.6  & 0.92 $\pm$ 0.02      \\

S28 & 06 56 40.1 ; ~+06 36 33.9   & 1.15 $\pm$ 0.03      \\
\hline
\end{tabular}
\end{table}

\subsection{Optical spectra}
Optical spectra were obtained at ten locations near the SW and NW regions of G206.7+5.9. The slit positions, labelled S30, S29, S28, S26, S23, S22, S21, S34, S53 and S52, are listed in Table~\ref{Table3}.

\begin{table}
 \caption{The summary of LAMOST observations.}
 \begin{tabular}{@{}p{1.2cm}p{3.4cm}p{1.6cm}@{}}
 \hline
Slit  &  R.A. (J2000) ; Dec. (J2000)  & Region \\
 ID & (h m s)  ~~~~~~~~~~~ ($\degr$ $\arcmin$ $\arcsec$)  & \\
\hline
S30 & 06 58 16.8 ; ~+07 17 44.6   &  SW \\
S29 & 06 57 33.5 ; ~+07 21 15.4   &  SW  \\
S28 & 06 56 40.1 ; ~+06 36 33.9   &  SW  \\
S26 & 06 56 43.6 ; ~+06 43 42.6   &  SW  \\
S23 & 06 57 05.3 ; ~+06 33 02.0   &  SW  \\
S22 & 06 57 30.4 ; ~+06 42 53.3   &  SW \\
S21 & 06 57 12.3 ; ~+06 47 19.4   &  SW  \\
S34 & 06 57 17.3 ; ~+06 57 08.3   &  SW  \\
S53 & 06 54 24.6 ; ~+09 29 28.3   &  NW  \\
S52 & 06 54 31.9 ; ~+09 25 54.7   &  NW  \\
\hline
\label{Table3}
\end{tabular}
\end{table}

We employed a multi-step fitting approach using Gaussian profiles to analyse the emission lines in the spectra of SNR regions observed by the LAMOST. The spectral data were first pre-processed to remove wave-like fluctuations by fitting and subtracting a polynomial of degree 1-5 depending on the fluctuations, isolating the emission lines for clearer analysis. The user was then prompted to visually select the approximate centers of the emission lines, which served as initial estimates for the Gaussian fitting.

The emission lines were modelled as a sum of Gaussian functions, where the number of Gaussians to be fitted (ranging from 1 to 3) was determined based on the complexity of the spectral region. Each Gaussian function was parameterised by its peak position (central wavelength), width (sigma), and amplitude (flux).

The fitting process was carried out using the curve$\_$fit function from the {\sc scipy.optimize}\footnote{\url{https://docs.scipy.org/doc/scipy/reference/optimize.html}} package, which applies a non-linear least-squares optimization to match the model to the detrended data. The initial parameters for each Gaussian (center, sigma, and amplitude) were derived from the user-selected peaks. This approach allowed for greater flexibility in fitting complex spectral features with multiple overlapping lines.

The uncertainties in the fitted parameters were quantified using the covariance matrix produced by the curve$\_$fit routine. The diagonal elements of the covariance matrix were used to compute the standard errors for each parameter (i.e., central wavelength, sigma, and flux). These uncertainties reflect the statistical errors associated with the fitting process and were reported alongside the best-fit values.

For each Gaussian component, the peak position, width, and flux were expressed as follows:

\begin{itemize}
\item Peak position (central wavelength): $x_0 \pm \sigma_{x_0}$
\item Width (sigma): $\sigma \pm \sigma_{\sigma}$
\item Amplitude (flux): A $\pm$ $\sigma_A$
\end{itemize}

These values provide both the physical parameters of the emission lines and their associated uncertainties, facilitating the interpretation of line strengths and profiles in the context of the SNR's physical conditions.

The LAMOST spectra exhibit several significant feature lines, including [O\,{\sc iii}]$\lambda$5007, [O\,{\sc i}]$\lambda$6300, $\lambda$6363, [N\,{\sc ii}]$\lambda$6548, H$\alpha$$\lambda$6563,  [N\,{\sc ii}]$\lambda$6584 and [S\,{\sc ii}]$\lambda$6716, $\lambda$6731. The H$\beta$$\lambda$4861 line is not observed because it falls outside the wavelength range. We presented the spectra in the 6540$-$6600 {\AA} and 6710$-$6750 {\AA} range in Figs~\ref{slits1}$-$\ref{slits3}. The spectra, including [O\,{\sc iii}]$\lambda$5007 and [O\,{\sc i}]$\lambda$6300, $\lambda$6363 emission lines, are shown in  Appendix~\ref{slits4}. Smoothing was performed for low signal/noise spectra and this was indicated in the related figures. The resulting emission-line measurements, line ratios ([S\,{\sc ii}]/H$\alpha$, [N\,{\sc ii}]/H$\alpha$, and [S\,{\sc ii}] $\lambda$6716/$\lambda$6731), and electron density ($n_{\rm e}$) with errors are shown in Table~\ref{Table4}.

\begin{figure*}
\includegraphics[angle=0, width=8.2cm]{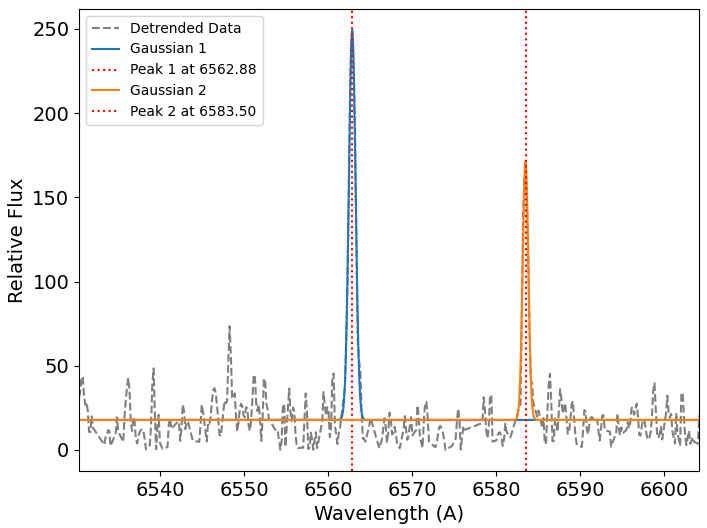}
\includegraphics[angle=0, width=8.2cm]{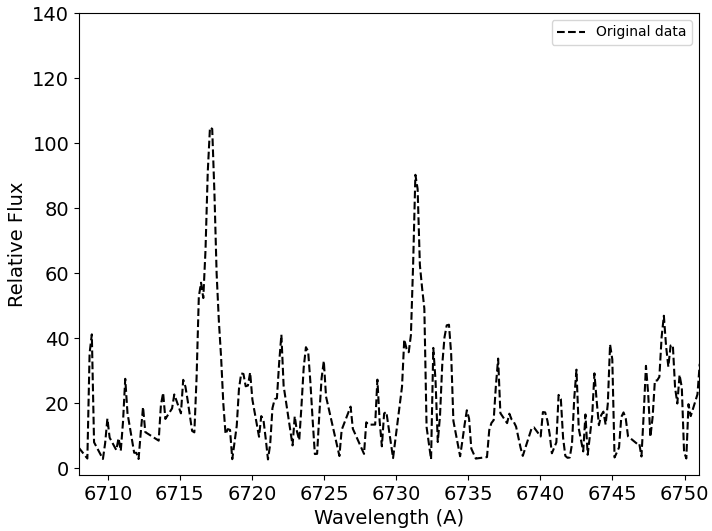}
\includegraphics[angle=0, width=8.2cm]{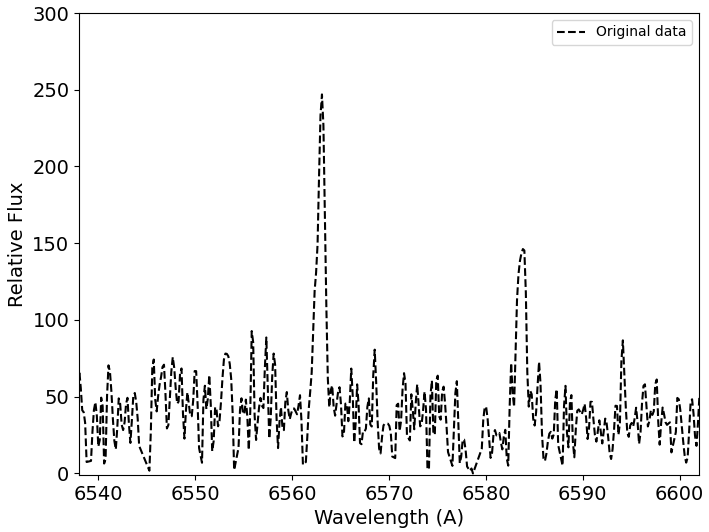}
\includegraphics[angle=0, width=8.2cm]{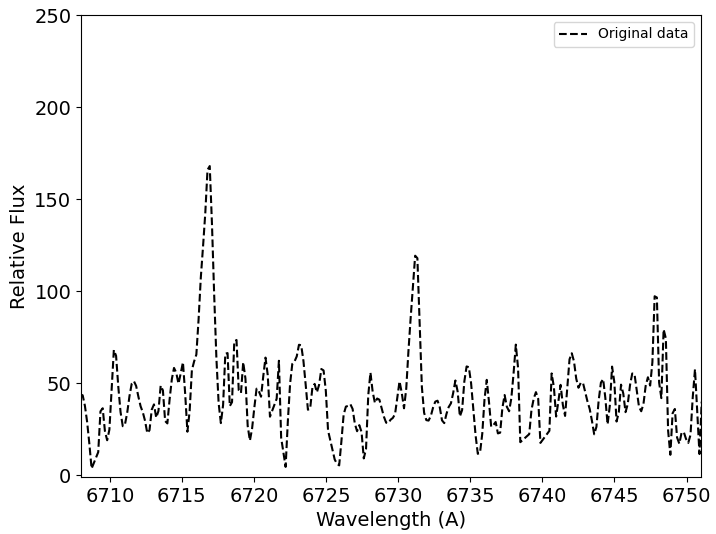}
\includegraphics[angle=0, width=8.2cm]{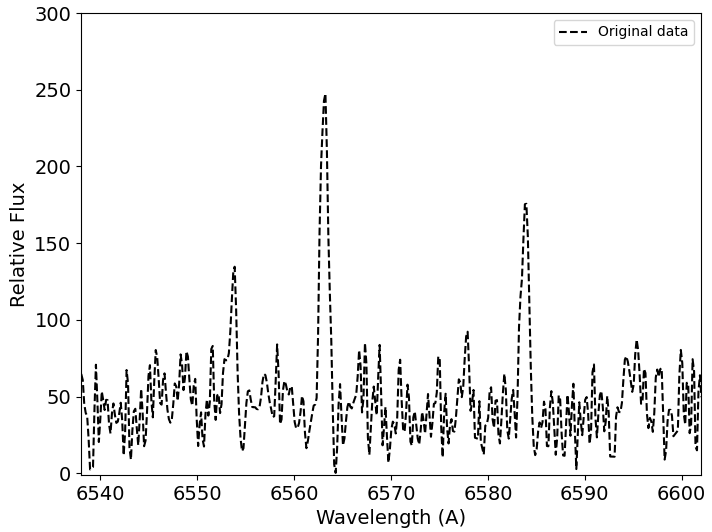}
\includegraphics[angle=0, width=8.2cm]{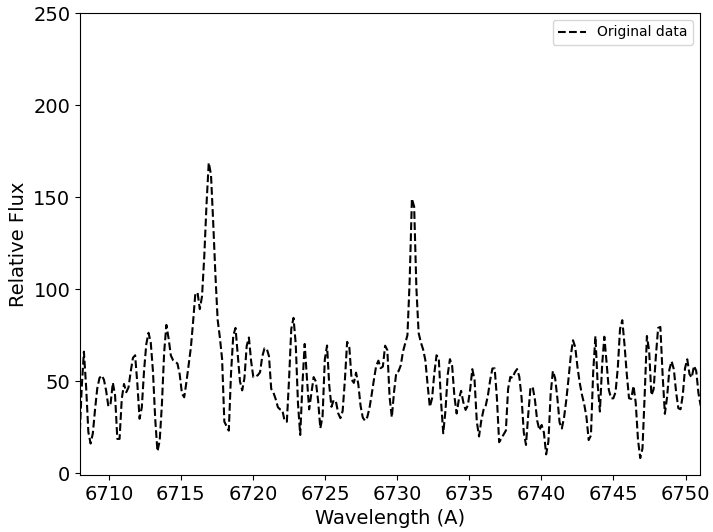}
\includegraphics[angle=0, width=8.2cm]{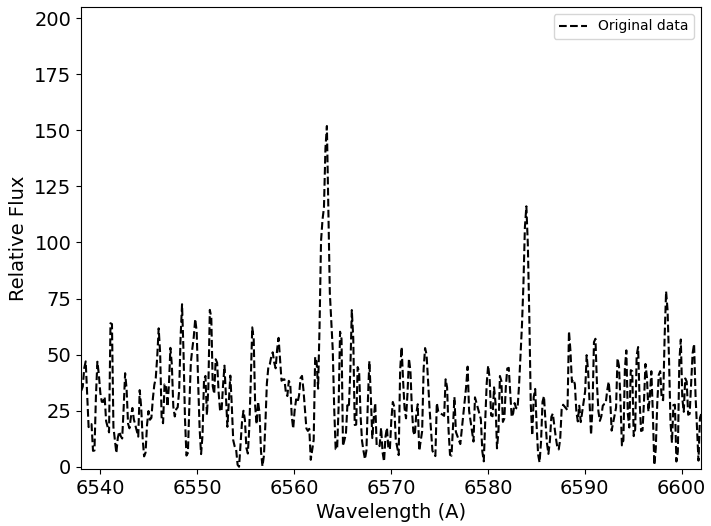}
\includegraphics[angle=0, width=8.2cm]{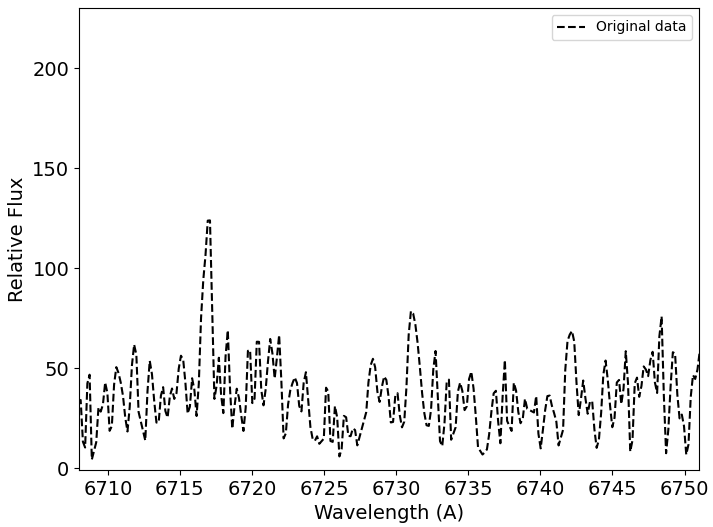}
\caption{The LAMOST spectra (6540$-$6600 {\AA} and 6710$-$6750 {\AA}) for the SW region (S28, S21, S22, and S23, respectively). Some relatively low S/N spectra are shown after smoothing, which are shown in the legend.}
\label{slits1}
\end{figure*}

\begin{figure*}
\includegraphics[angle=0, width=8.2cm]{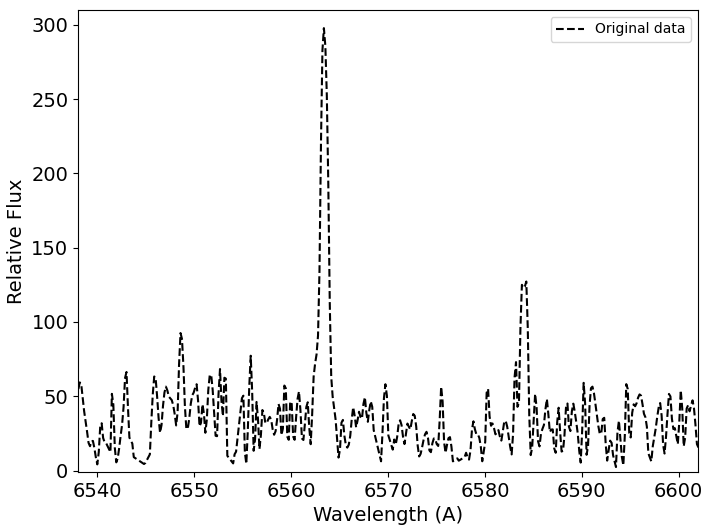}
\includegraphics[angle=0, width=8.2cm]{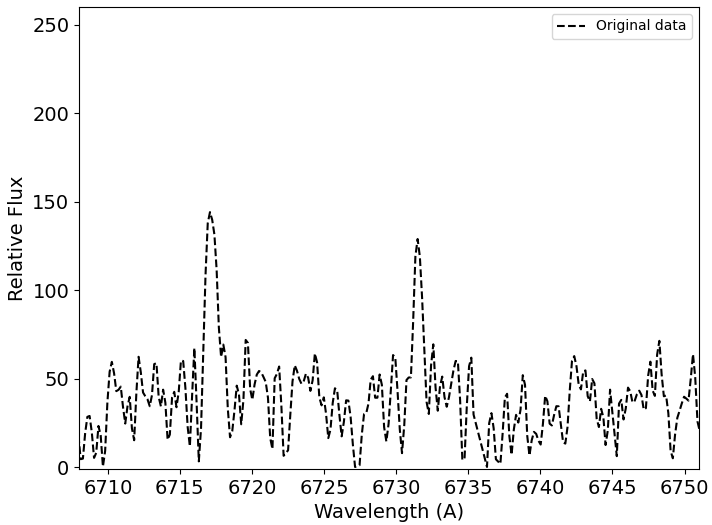}
\includegraphics[angle=0, width=8.2cm]{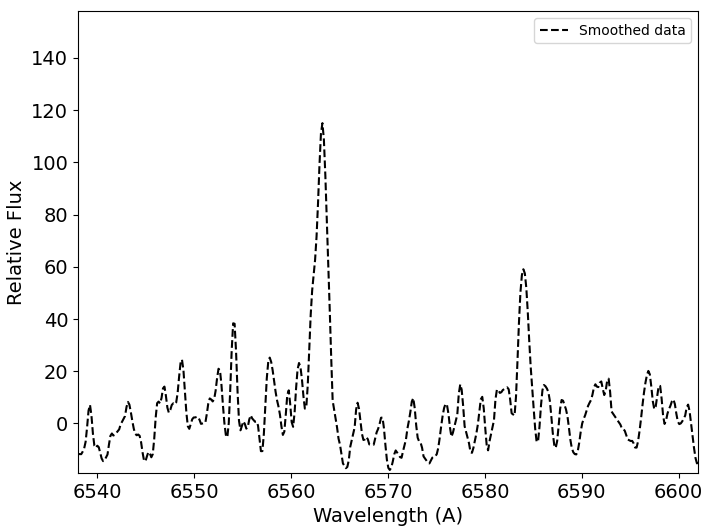}
\includegraphics[angle=0, width=8.2cm]{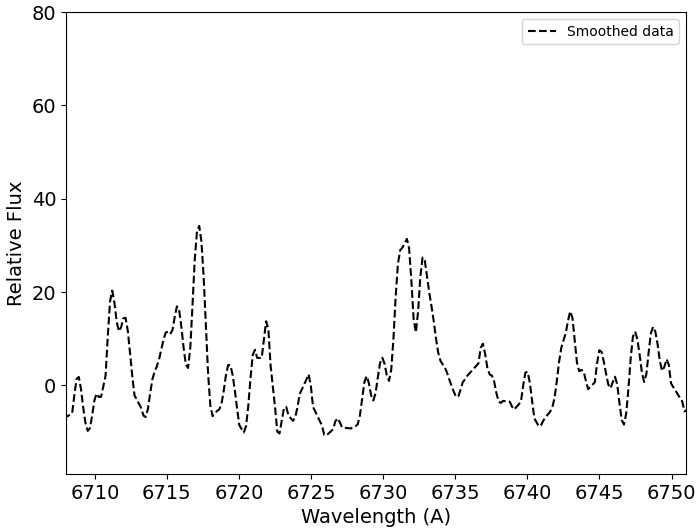}
\includegraphics[angle=0, width=8.2cm]{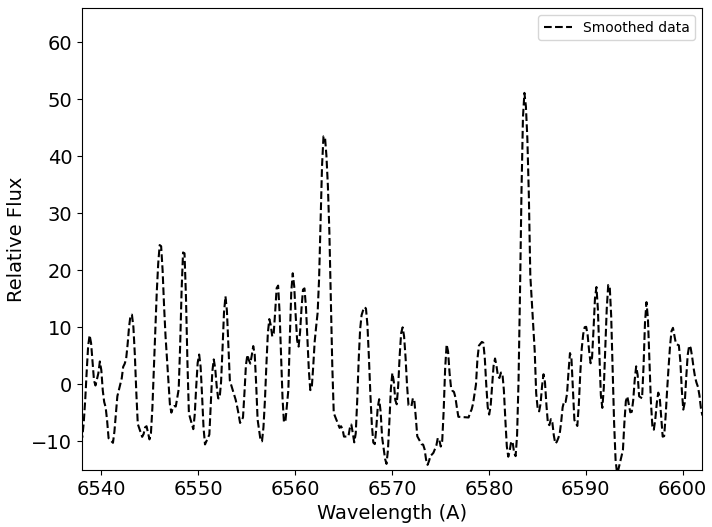}
\includegraphics[angle=0, width=8.2cm]{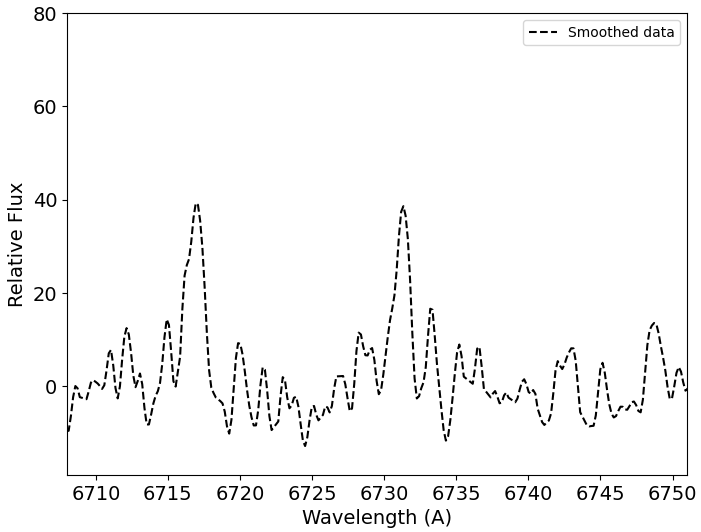}
\includegraphics[angle=0, width=8.2cm]{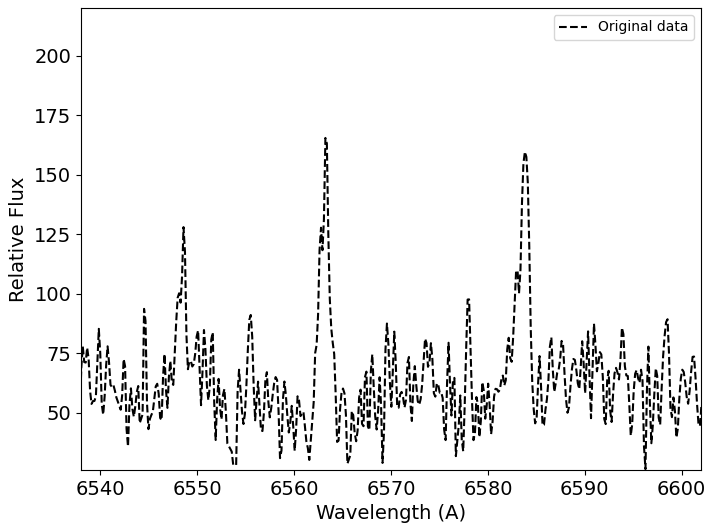}
\includegraphics[angle=0, width=8.2cm]{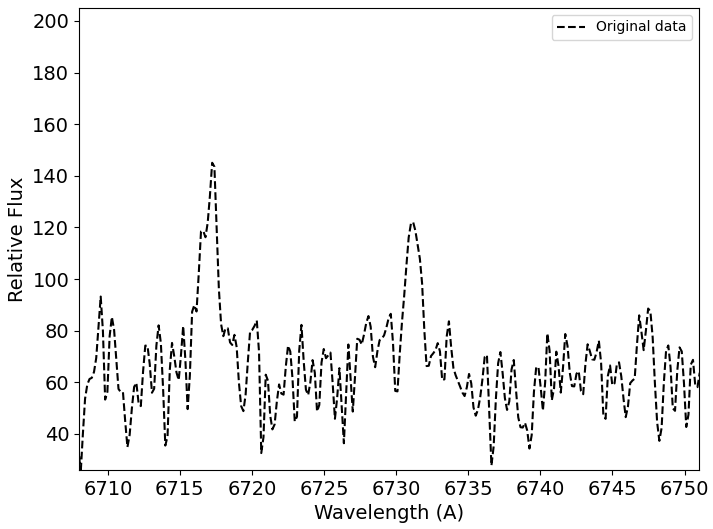}
\caption{Continued from Fig.~\ref{slits1}. Spectra for the SW region (S26, S29, S30, and S34, respectively).}
\label{slits2}
\end{figure*}

\begin{figure*}
\includegraphics[angle=0, width=8.2cm]{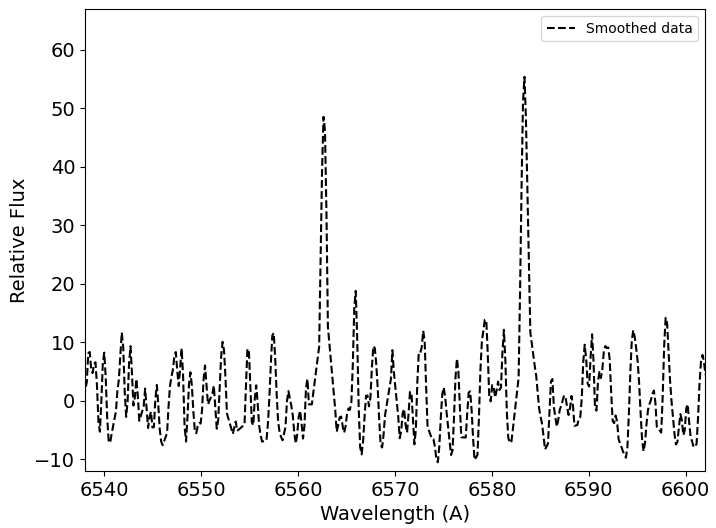}
\includegraphics[angle=0, width=8.2cm]{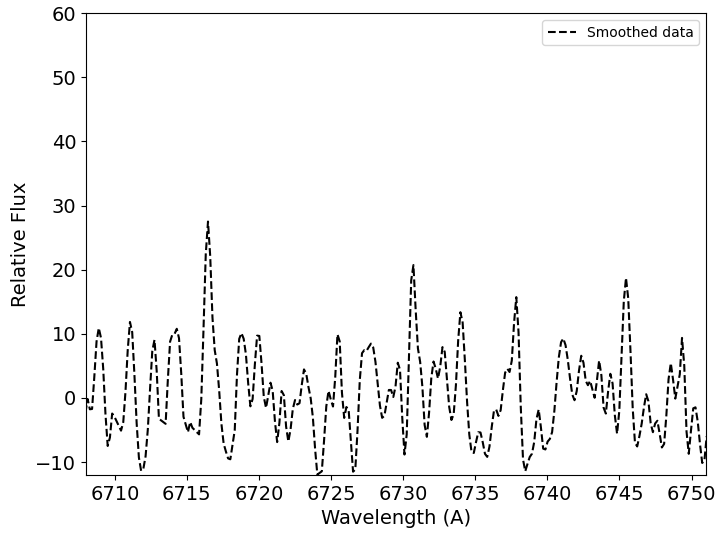}
\includegraphics[angle=0, width=8.2cm]{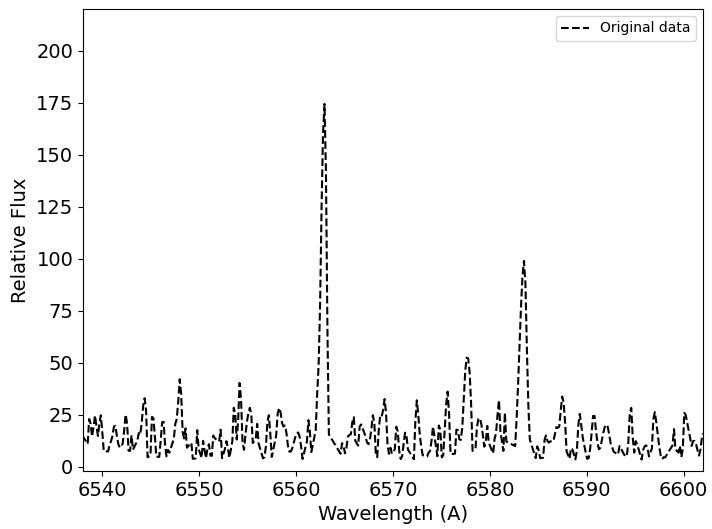}
\includegraphics[angle=0, width=8.2cm]{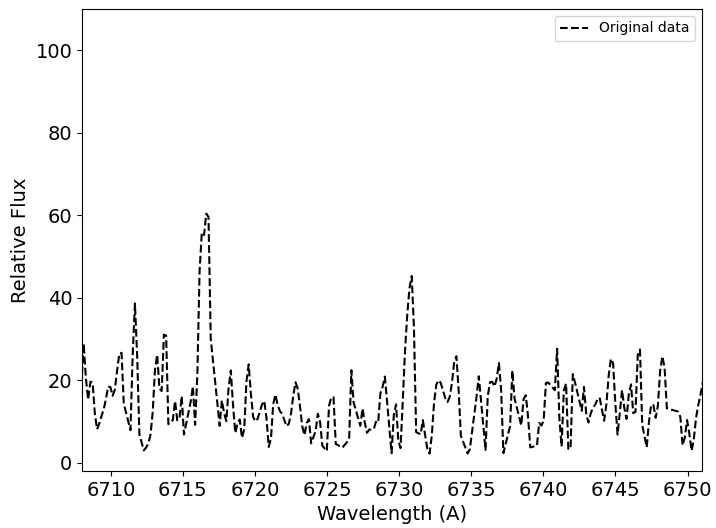}
\caption{Continued from Fig.~\ref{slits1}. Spectra for the NW region (S52 and S53).}
\label{slits3}
\end{figure*}

\begin{table*}
\centering
\caption{Relative line fluxes are given on a scale where $F$(H$\alpha$) = 100 with the 1-$\sigma$ errors. The line ratios and electron densities are also presented.}
\label{Table4}
 \begin{tabular}{@{}p{3.5cm}p{1.8cm}p{1.8cm}p{1.8cm}p{1.8cm}p{1.8cm}p{1.8cm}@{}}
 \hline
  &	S30	&	S29	&	S28 &	S26 &	S23		 \\[0.5 ex]
\hline		

$[$O$\,${\sc iii}$]$ ($\lambda$5007) 	&	62$\pm$7	&	$...$	&   20$\pm$5	&	10$\pm$1 &	$...$	  \\
												
$[$O$\,${\sc i}$]$ ($\lambda$6300) 	& $...$  &	40$\pm$2	&	$...$ &	$...$ &	$...$		 \\
												
$[$O$\,${\sc i}$]$ ($\lambda$6363) 	&	$...$	       &	$...$	&	$...$ &	$...$	&	$...$	 \\
												
$[$N$\,${\sc ii}$]$ ($\lambda$6548) 	&	69$\pm$6	&	41$\pm$1	&	22$\pm$2 &	31$\pm$1	&	$...$	  \\

H$\alpha$ ($\lambda$6563) 	&	100$\pm$8	&	100$\pm$5	&	100$\pm$2 &	100$\pm$4	&	100$\pm$5	  \\
												
$[$N$\,${\sc ii}$]$ ($\lambda$6584) 	&	123$\pm$12	&	65$\pm$3 	&	68$\pm$4 &	42$\pm$4 &	79$\pm$5	  \\
												
$[$S$\,${\sc ii}$]$ ($\lambda$6716) 	&	92$\pm$8	&	41$\pm$5	&	38$\pm$2 &	49$\pm$3 &	89$\pm$6    \\
												
$[$S$\,${\sc ii}$]$ ($\lambda$6731) 	&	86$\pm$8	&	40$\pm$4	& 30$\pm$3 &	42$\pm$2	 &	54$\pm$3	\\[0.5 ex]
												
[S\,{\sc ii}] ($\lambda$6716+$\lambda$6731)/ H$\alpha$ 	&	  1.78$\pm$0.02  &	  0.80$\pm$0.05 	&	   0.69$\pm$0.02 &	0.91$\pm$0.02 &	1.42$\pm$0.04	  \\[0.5 ex]

[N\,{\sc ii}] ($\lambda$6548+$\lambda$6584)/ H$\alpha$ 	&	   1.92$\pm$0.02  &	  1.06$\pm$0.01 	&	   0.90$\pm$0.04 &	0.73$\pm$0.02 &	0.79$\pm$0.01	 \\[0.5 ex]

[S\,{\sc ii}] $\lambda$6716/$\lambda$6731 	&	1.08$\pm$0.01    &	   1.03$\pm$0.01 	&	 1.27$\pm$0.06 &	1.17$\pm$0.02 &	1.66$\pm$0.03	 \\[0.5 ex]
												
Electron density $n_{\rm e}$(cm$^{-3}$)	&	   509$\pm$8 	&	  597$\pm$30	&	     204$\pm$26 &	338$\pm$18 &	$...$	 \\
\\[0.5 ex]	
\hline
 &	S22	&	S21	&		S34 &		S53 &		S52	 \\[0.5 ex]
\hline												
            
$[$O$\,${\sc iii}$]$ ($\lambda$5007) 	&	$...$	 &	$...$		&	$...$ &	$...$	&  $...$  \\
												
$[$O$\,${\sc i}$]$ ($\lambda$6300) 	   & 35$\pm$1    &	$...$ &		96$\pm$1 &	46$\pm$2	& $...$ \\
												
$[$O$\,${\sc i}$]$ ($\lambda$6363) 	   &	$...$	  &	$...$	&		$...$	&	60$\pm$1 & $...$ \\
										
$[$N$\,${\sc ii}$]$ ($\lambda$6548) 	&	 $...$	    &	$...$	    &	57$\pm$8	& 19$\pm$2	& $...$  \\

H$\alpha$ ($\lambda$6563) 	&	100$\pm$4	&	100$\pm$5	&		100$\pm$11	&	100$\pm$5 &	100$\pm$2   \\
																	
$[$N$\,${\sc ii}$]$ ($\lambda$6584) 	&	70$\pm$3	&	63$\pm$4 	 &	106$\pm$8 &	56$\pm$2	& 137$\pm$4  \\
												
$[$S$\,${\sc ii}$]$ ($\lambda$6716) 	&	53$\pm$3	&	67$\pm$3  &	83$\pm$8 &	35$\pm$3	& 59$\pm$7   \\
												
$[$S$\,${\sc ii}$]$ ($\lambda$6731) 	&	43$\pm$3	&	50$\pm$2	&	72$\pm$7	 &	26$\pm$2  &  53$\pm$8 \\[0.5 ex]
												
[S\,{\sc ii}] ($\lambda$6716+$\lambda$6731)/ H$\alpha$ 	&	  0.97$\pm$0.03  &	  1.17$\pm$0.01  &	1.55$\pm$0.02 &	0.61$\pm$0.03	&  1.13$\pm$0.11 \\[0.5 ex]

[N\,{\sc ii}] ($\lambda$6548+$\lambda$6584)/ H$\alpha$ 	&	   0.70$\pm$0.01  &	   0.63$\pm$0.01 	&		1.63$\pm$0.02 &	0.75$\pm$0.01 &  1.37$\pm$0.01 \\[0.5 ex]

[S\,{\sc ii}] $\lambda$6716/$\lambda$6731 	&	   1.24$\pm$0.02  &	   1.34$\pm$0.01 	&	1.15$\pm$0.01 &	1.35$\pm$0.01	& 1.11$\pm$0.04\\[0.5 ex]
												
Electron density $n_{\rm e}$(cm$^{-3}$)	&	    243$\pm$10 	&	  126$\pm$1	&	  	369$\pm$6 &	117$\pm$1 & 437$\pm$82	 \\
\\[0.5 ex]	
  \hline
\end{tabular}
\end{table*}

\subsection{Interstellar gaseous environment}
\begin{figure*}
\includegraphics[angle=0, width=\textwidth]{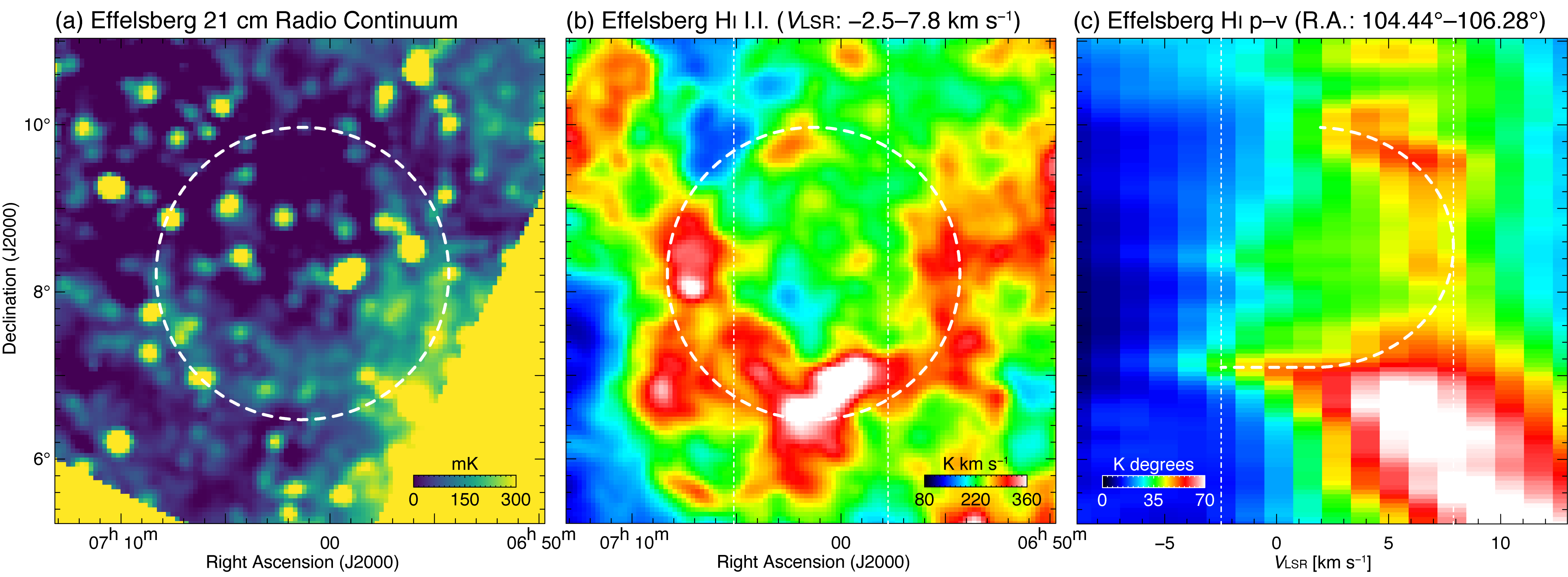}
\caption{(a) Distribution of radio continuum at 21 cm obtained using the Effelsberg 100-m radio telescope \citep{1999A&AS..138...31U}. The dashed circle indicates the shell boundary of the SNR G206.7+5.9. (b) Integrated intensity map of H{\sc i} obtained using the Effelsberg 100-m radio telescope \citep{2016A&A...585A..41W}. The integrated velocity range is from $-2.5$ to 7.8~km~s$^{-1}$. The superposed dashed circle is the same as shown in (a). (c) Position--velocity diagram of H{\sc i}. The integration R.A. range is from 104\fdg44 to 106\fdg28. The dashed curve indicates an expanding gas motion (see the text).}
\label{hi-results}
\end{figure*}

Figs \ref{hi-results}(a) and \ref{hi-results}(b) show maps of the 21 cm radio continuum and H{\sc i} line toward the SNR G206.7+5.9, respectively. The faint arc-like feature of the radio continuum and the obvious H{\sc i} shell are seen, which are roughly consistent with \citet{Ga22}. Fig.~\ref{hi-results}(c) shows the position--velocity (p--v) diagram of H{\sc i}. We newly identified an expanding gas motion of the H{\sc i}, whose velocity span is approximately 10 km s$^{-1}$. Since the Declination range of the H{\sc i} cavity in the p--v diagram is roughly the same as the shell diameter of the SNR, the expanding gas motion of the H{\sc i} was formed by strong stellar winds from the massive progenitor and/or supernova blast waves \citep[e.g.,][]{1990ApJ...364..178K,1991ApJ...382..204K}. In any case, our results are consistent with the scenario by \citet{Ga22}.

\section{Discussion}
\label{discuss}
In the following, we present the optical properties of G206.7+5.9 based on our new findings, compare G206.7+5.9 with other large size SNRs, and provide our conclusions regarding its nature.
\subsection{Optical properties of G206.7+5.9}
We observed the SW region of G206.7+5.9 using the H$\alpha$ and [S\,{\sc ii}] filters. The emission from this region shows filamentary structure seen in the H$\alpha$ image (Fig.~\ref{figure1b}, left panel). The SHASSA and VTSS images show that there is H$\alpha$ emission in a wide area around the remnant. We found a strong correlation between optical and radio emissions along the SNR, indicating that the emissions originate from the SNR shock (see Fig.~\ref{figure1a}, upper-right panel). The [S\,{\sc ii}]/H$\alpha$ ratio is the most conventional method used for the optical identification of SNRs (e.g. \citealt{Fe85}). Based on the CCD imaging, we found an average [S\,{\sc ii}]/H$\alpha$ ratio of approximately 0.74, indicating that the emission primarily originates from shock-heated gas.

Spectra from G206.7+5.9 reveal abundant emission lines (Figs \ref{slits1}$-$\ref{slits3}). We calculated some line ratios, as shown in Table~\ref{Table4}, since the physical parameters of the ionized gas can be determined using the ratios of certain emission lines \citep{OsFe06}. For example; the ratio of the [S\,{\sc ii}]$\lambda$6716, $\lambda$6731 lines can be used to estimate the electron density ($n_{\rm e}$) and is nearly independent of electron temperature ($T$) \citep{OsFe06}. From the [S\,{\sc ii}]$\lambda$6716/$\lambda$6731 lines ratio and assuming electron temperature $T$=$10^{4}$ K, we derived the density of the region of G206.7+5.9. The electron densities between 117$-$597 cm$^{-3}$ with an average ratio around 327 cm$^{-3}$. For the slit position (S53), we measured a [S\,{\sc ii}]$\lambda$6716, $\lambda$6731 ratio of approximately 1.35, indicating a density close to the low-density limit ($n_{\rm e}$ $\sim$ 117 cm$^{-3}$). In contrast, the ratio for S29 was estimated to be $\sim$1.03, suggesting a much higher electron density ($\sim$597 cm$^{-3}$) than at the other slit locations.

The [S\,{\sc ii}]/H$\alpha$ line ratios from the LAMOST spectra ($\sim$0.61$-$1.78; see Table~\ref{Table4}), which are greater than 0.4$-$0.5, indicate the presence of shocks. The [N\,{\sc ii}]/H$\alpha$ line ratios are also generally high (0.63$-$1.92). Both ratios rule out any possible confusion with H\,{\sc ii} regions.

The detection of [O\,{\sc i}] $\lambda$6300, $\lambda$6363 emission lines in the LAMOST spectra further supports the presence of shock activity. The [O\,{\sc i}] $\lambda$6300, $\lambda$6363 lines can be affected by imperfect subtraction of the strong [O\,{\sc i}] night sky emission. Principal component analysis (PCA) is a well-established technique that has been applied to sky subtraction in fiber spectroscopy \citep{Bai17}. Although LAMOST's PCA-based sky subtraction reduces the average residuals, some emission lines may still be contaminated by residuals (e.g. \citealt{Seok20}). Nevertheless, some spectra in our dataset  which are obtained in the same observing night do not show any signs of [O\,{\sc i}] line suggesting a non-existent or very weak contribution of the night sky emission in this wavelength.

We calculated the pre-shock cloud density $n_{\rm c}$ $\sim$2.6$-$13.3 cm$^{-3}$, assuming a shock velocity $V_{\rm s}$ $\sim$ 100~km~s$^{-1}$, using the relation 

\begin{equation}
n_{\rm [S\,{\sc II}]} = 45~n_{\rm c} \times (V/100~{\rm km~s^{-1}})^{2}~~~{\rm cm}^{-3}
\end{equation}

from \citet{Do79, Fe80}, where $n_{\rm [S\,{\sc II}]}$ is the electron density (see Table~\ref{Table4}). The range of pre-shock cloud densities suggests that the SNR is interacting with a high-density clumpy ISM.

The age of the SNR in the cooling phase ($t_{\rm cool}$) can be estimated using the relation from \citet{Fa81}
\begin{equation}
    t_{\rm cool} = 2.7 \times 10^{4} ~ E_{51}^{0.24} ~ n_{\rm 0}^{-0.52} ~ ~ {\rm yr},
\end{equation}
where $E_{51}$ represents the average explosion energy in units of $10^{51}$~erg and $n_{\rm 0}$ denotes the pre-shock density in cm$^{-3}$. By adopting $E_{51}$ = 0.5 and pre-shock densities of $\sim$2.6$-$13.3 cm$^{-3}$, we estimated the range of the SNR age to be $\sim$(0.6$-$1.4) $\times$ 10$^{4}$ years.

The absorption column density ($N_{\rm H}$) obtained from the \texttt {nh} tool\footnote{\url{https://heasarc.gsfc.nasa.gov/cgi-bin/Tools/w3nh/w3nh.pl}} is  $\sim$1.67$\times10^{21}$~cm$^{-2}$ \citep{Di90}. Using Eq. (9) from \citet{Fo16}, we converted the $N_{\rm H}$ value to an optical extinction ($A_{\rm V}$), resulting in a value of approximately 0.58. Subsequently, using optical extinction datasets\footnote{\url{https://explore-platform.eu/sda/g-tomo/}} from \citet{La22}, we derived an $A_{\rm V}$ of $\sim$0.1 at a distance $\sim$0.44 kpc given by \citet{Ga22}. The difference between the two $A_{\rm V}$ values obtained using different methods suggests that the distance of SNR exceeds 0.44 kpc.

\subsection{Comparisons of G206.7+5.9 and other large-size shell type SNRs}
We compared G206.7+5.9 with other larger angular size shell-type Galactic SNRs (from Green's catalog\footnote{\url{https://www.mrao.cam.ac.uk/surveys/snrs/}})  showing optical emission. 

There are some remnants similar to G206.7+5.9. For example, G156.2+5.7 is a large ($110 \times 110$ arcmin$^{2}$) SNR, and its H$\alpha$ image reveals numerous non-radiative filaments \citep{Ka16}. The optical spectra of the SNR indicate moderately high post-shock densities ($n_{\rm e}$) of around 200$-$300 cm$^{-3}$, a pre-shock density ($n_{\rm c}$) of approximately 10 cm$^{-3}$, and a [S\,{\sc ii}]/H$\alpha$ ratio greater than 0.4 \citep{Ge07}. 

For the S30, S34, and S52 slit positions, we measured a strong [N\,{\sc ii}]$\lambda$6584 line (see Table~\ref{Table4}). A similar feature was observed in optical spectroscopic studies of other large shell-type SNRs, such as G279.0+1.1 ($95 \times 95$ arcmin$^{2}$) \citep{St09} and G315.1+2.7 ($190 \times 150$ arcmin$^{2}$) \citep{St07}. According to \citet{Ki80}, the observed nitrogen strength is primarily due to the interaction of SNRs with the enriched material in the ISM, which they sweep up, rather than originating from the supernova ejecta itself.

\section{Conclusions}
\label{conc}
In this work, we have investigated the optical properties of the SW and NW regions of G206.7+5.9.
The H$\alpha$ and [S\,{\sc ii}] narrow-band images of the SW region, taken with the 1-m T100 telescope, show filamentary structure. The LAMOST spectra reveal large ratios of [S\,{\sc ii}]/H$\alpha$ $\sim$ (0.61$-$1.78) and [N\,{\sc ii}]/H$\alpha$ $\sim$ (0.63$-$1.92) consistent with that expected for a shock-heated SNR. The [O\,{\sc i}] $\lambda$6300, $\lambda$6363 line detected in the spectra also support the presence of shocks. The [S\,{\sc ii}] $\lambda$6716/$\lambda$6731 ratios indicate densities ranging from 117 up to 597 cm$^{-3}$.  We concluded that G206.7+5.9 is an SNR with characteristics strikingly similar to those observed in Galactic SNRs.

Future, more detailed multifrequency studies will significantly enhance our understanding of the physical processes in this large Galactic SNR G206.7+5.9. These findings contribute to a more detailed understanding of SNR evolution and their interaction with the ISM.

\section*{Acknowledgements}
We thank the referee for valuable comments and suggestions that helped to improve the paper. We also thank T\"{U}B\.{I}TAK National Observatory for partial support in using T100 telescope with project number 2153. This research was supported by the Scientific and Technological Research Council of T\"{u}rkiye (T\"{U}B\.{I}TAK) through project number 124F089. This work has made use of data products from the Guoshoujing Telescope (the Large Sky Area Multi-Object Fiber Spectroscopic Telescope, LAMOST). LAMOST is a National Major Scientific Project built by the Chinese Academy of Sciences. Funding for the project has been provided by the National Development and Reform Commission. LAMOST is operated and managed by the National Astronomical Observatories, Chinese Academy of Sciences. 
This work has used the image obtained by the the Virginia Tech Spectral-Line Survey (VTSS), which is supported by the National Science Foundation and by the Southern H-Alpha Sky Survey Atlas (SHASSA), which is supported by the National Science Foundation. This work also has used tools developed as part of the EXPLORE project that has received funding from the European Union’s Horizon 2020 research and innovation program under grant agreement No. 101004214. This work was supported by JSPS KAKENHI grant No. 21H01136 (HS), 24H00246 (HS).

\section*{DATA AVAILABILITY}
The optical data from the T100 telescope will be shared on reasonable request to the corresponding author. The LAMOST data can be accessed at https://www.lamost.org/dr9/v2.0/.

 

\clearpage

\bibliographystyle{mnras}
\bibliography{example} 



\appendix
\section{ADDITIONAL PLOTS}
\begin{figure*}
\includegraphics[angle=0, width=8.2cm]{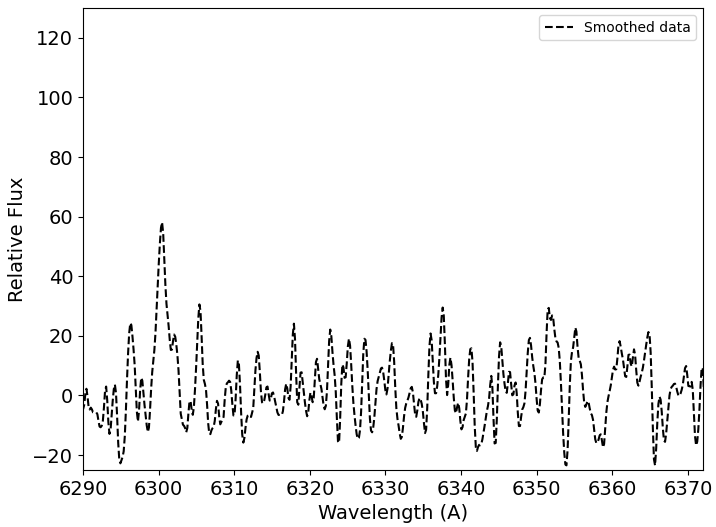}
\includegraphics[angle=0, width=8.2cm]{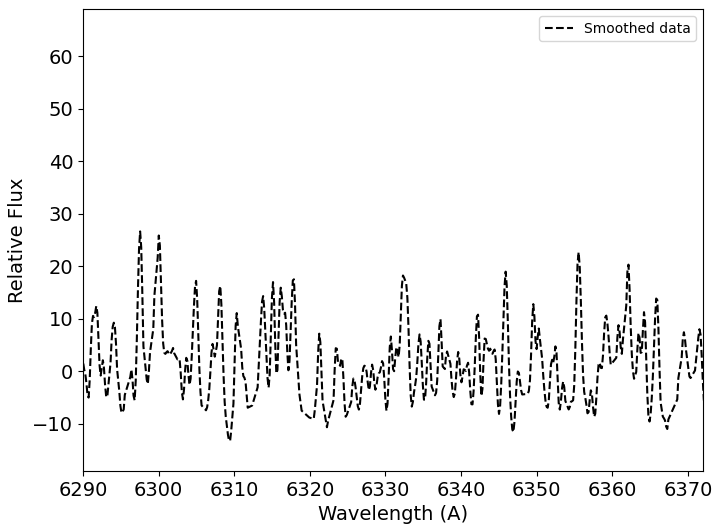}
\includegraphics[angle=0, width=8.2cm]{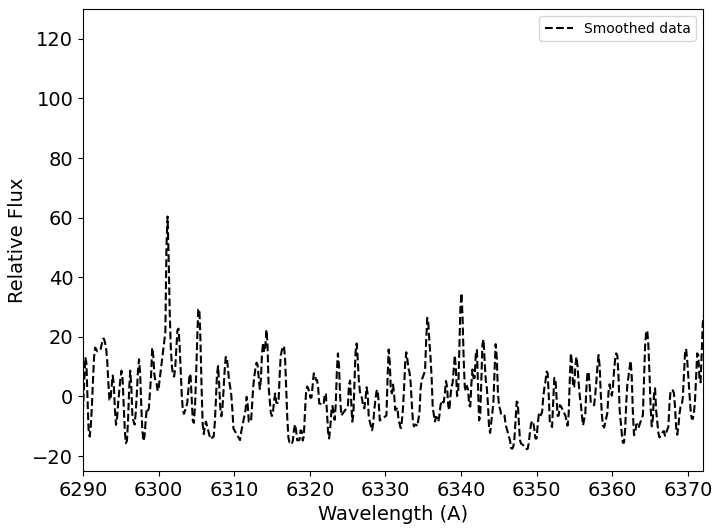}
\includegraphics[angle=0, width=8.2cm]{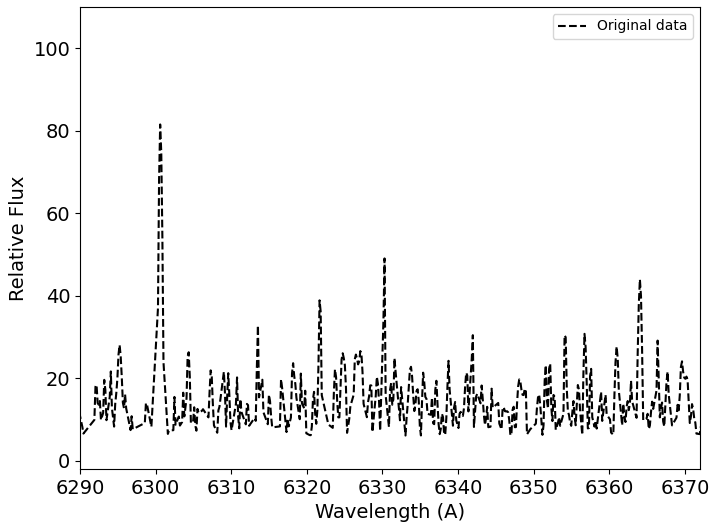}
\includegraphics[angle=0, width=8.2cm]{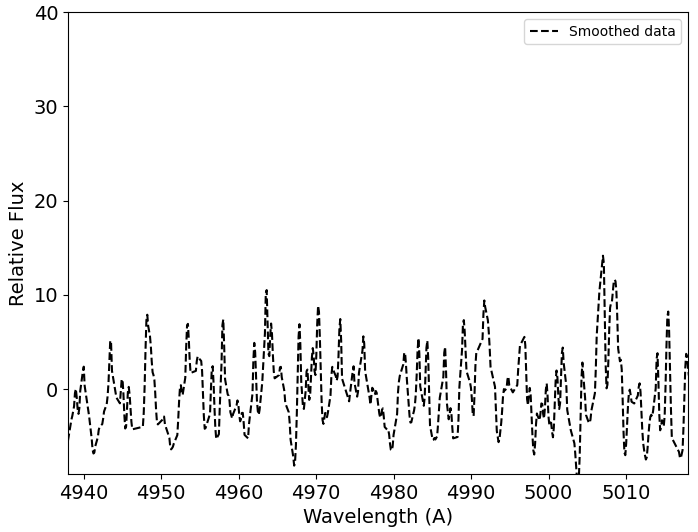}
\includegraphics[angle=0, width=8.2cm]{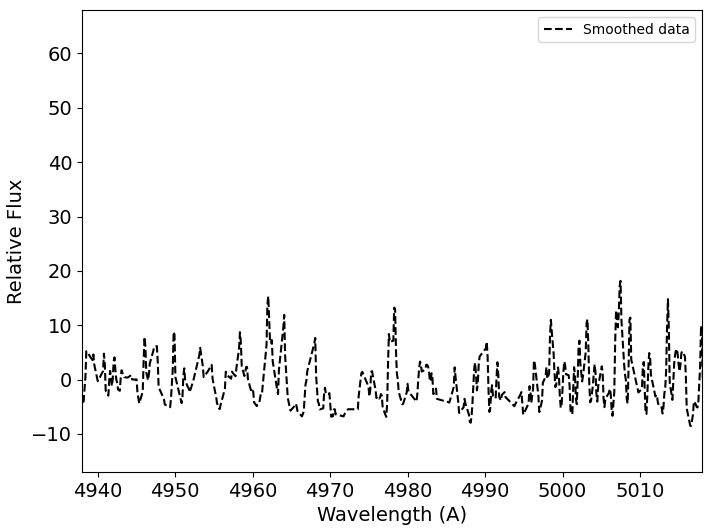}
\includegraphics[angle=0, width=8.2cm]{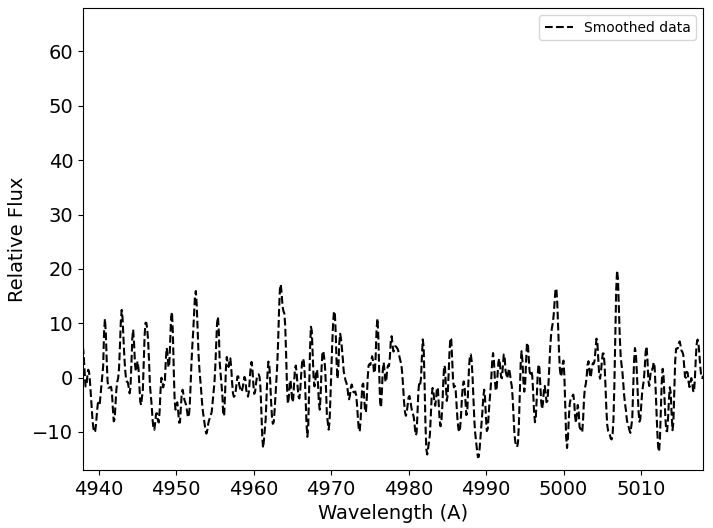}

\caption{LAMOST spectra cover wavelengths from 6290 to 6370 {\AA} for S22, S29, S34, and S53 and from 4940 to 5010 {\AA} for S28, 26, and S30, respectively.}
\label{slits4}
\end{figure*}

In Fig.~\ref{slits4}, we show LAMOST spectra (4940$-$5010 {\AA} and 6290$-$6370 {\AA}), including [O\,{\sc iii}]$\lambda$5007 and [O\,{\sc i}]$\lambda$6300, $\lambda$6363 emission lines.

\bsp	
\label{lastpage}
\end{document}